\documentstyle[twocolumn]{article}
\def\kmsmpc{${\rm ~km ~s^{-1}Mpc^{-1}}$ }
\def\kms-1{~{\rm km~s^{-1}}}
\def\kms{km s$^{-1}$ }

\def\Msun{M_\odot}
\def\re{\hangindent 1pc \noindent}
\begin{document}
\textwidth=16cm
\textheight=24cm
\oddsidemargin=0pt
\evensidemargin=0pt
\title{Hubble Constant at Intermediate Redshift using the
CO-Line Tully-Fisher Relation}
\author{
Yoshinori {\sc Tutui},$^{1, 2}$
Yoshiaki {\sc Sofue},$^1$
Mareki {\sc Honma},$^{3}$
Takashi {\sc Ichikawa},$^{4}$ \\
and Ken-ichi {\sc Wakamatsu}$^5$
\\
$^1$ {\it Institute of Astronomy, University of Tokyo,
        Mitaka, Tokyo 181-8588, Japan}\\
{\it Email: sofue@ioa.s.u-tokyo.ac.jp}\\
$^2$ {\it NHK, Science Division, Jin-nan, Shibuya-ku, Tokyo 150-0041, Japan}\\ 
$^3$ {\it VERA Project Office, National Astronomical Obs. Japan
        Mitaka, Tokyo 181-8588, Japan}\\ 
$^4$ {\it Astronomical Institute, Tohoku University,
        Aoba, Sendai 980-8578, Japan}\\
$^5$ {\it Department of Technology, Gifu University,
        1-1 Yanagido, Gifu 501-11, Japan}
}
\maketitle
\begin{abstract}
We have determined distances and Hubble ratios for galaxies at
intermediate redshifts, $cz \sim 10,000$ to 35,000 \kms, by
applying the CO-line Tully-Fisher relation to our
$^{12}$CO $(J=1-0)$-line observations using the Nobeyama  45-m telescope, 
and near-IR (NIR) photometry in the $J$ and $H$ bands using the 1.88-m 
telescope at  Okayama Astrophysical Observatory.
By averaging the Hubble ratios from J-band result, 
we obtain a Hubble constant of $H_0 = 60 \pm 10$ \kmsmpc.
We argue that the CO line-NIR TFR can be a complimentary method to
the other methods for measuring distances of galaxies at intermediate 
and high redshifts.

\end{abstract}
%%%%%%%%%%%%%%%%%%%%%%%%%%%%%%%%%
%  Introduction
%%%%%%%%%%%%%%%%%%%%%%%%%%%%%%%%%
\section{Introduction}

The HI-line Tully-Fisher relation (TFR) has been established as one of the
most reliable methods to determine the distances of galaxies,
and has been successfully applied for   detemining the Hubble constant at 
(Tully and Fisher 1977; Aaronson et al 1980;  Giovanelli et al 1985, 1995; 
Haynes and Giovanelli 1986).
The use of the CO lines has been proposed as a complimentary method
to HI (Sofue 1992; Dickey \& Kazes 1992; Sofue et al 1996).
Compatibility between CO and HI linewidths has been confirmed for nearby galaxy
samples (Sch\"oniger \& Sofue 1994, 1997; Lavezzi and Dickey 1998;
Tutui \& Sofue 1999; Tutui et al 2000).

The CO-line TFR has some advantages, particluarly at higher redshifts:
Since the molecular-gas distribution in a galaxy is tightly
correlated with the luminousity distribution, CO-line width
can be properly compared with photometric luminosity.
CO gas distributions are less affected by galaxy interactions
and intracluster medium compared to HI, and the CO gas is not particularly 
defficient even in cluster-center galaxies (Kenney \& Young 1988; Casoli 
et al. 1991), which enables us to apply the TFR to rich-cluster galaxies.
These properties will be helpful to apply the TFR to high redshifts,
at which the fraction of interacting and merging galaxies is supposed to be 
greater than at low redshifts. 
Since high-redshift galaxies are supposed to be more dusty than nearby 
galaxies, near-IR photometry will be more appropriate than optical. 
Thus, the CO-NIR TFR would be a promising method to determine distances
of intermediate and high-redshift galaxies, and particularly for cluster 
galaxies at high redshifts.

In this paper we report the result of an extensive CO TFR program using the
Nobeyama 45-m mm-wave telescope, and NIR photometry using
a 1.88-m telescope at Okayama Observatory.
The program has been performed as a pilot program to promote a
more sensitive, higher resolution CO-NIR TFR project using the Nobeyama
mm-wave Array, and an advanced array linked to the 45-m telescope.
We obviously aim at obtaining basic data for higher-transition
CO line TFR using the Atacama Large Mm-wave and sub-mm wave Array (ALMA) 
in the near future. 

%%%%%%%%%%%%%%%%%%%%%%%%%%%%%%%%
%  CO observations
%%%%%%%%%%%%%%%%%%%%%%%%%%%%%%%%

\section{Observations}

\subsection{Sample selection}

We selected isolated galaxies of normal morphology
as a sample according to the following criteria.

(1) Redshift range of the sample was taken to be
$cz$  $ = 10,000 - 20,000 ~{\rm km~s^{-1}}$ in 1994/1995 observations,
$cz$  $ = 20,000 - 30,000 ~{\rm km~s^{-1}}$ in 1995/1996,
and   $cz$  $ = 30,000 - 50,000 ~{\rm km~s^{-1}}$
in 1996/1997 observations. The CO data are published in Tuttui
et al (2000) for discussion of CO and infrared properties of
non-interacting IRAS galaxies at intermediatet redshifts.

(2) We selected relatively strong FIR-emission sources
at 60 $\mu\rm m$ and 100 $\mu\rm m$, typically greater than 1 Jy,
which are supposed to be bright in the CO line.
Such FIR luminous galaxies are supposed to be affetected by
efficient star formaiton. For this reason, we use NIR luminosities,
istead of blue luminosities, which may be strongly affected by the
star formation.
However, we also emphasize that the TFR applies to IRAS galaxies
with much higher FIR luminosities (van Driel et al 1995).

(3) In order to minimize the effect of galaxy-galaxy interaction,
we selected galaxies with normal morphology using
DSS (STScI Digitized Sky Survey) images. Strongly interacting galaxies
and mergers were not included. They are not likely be starburst galaxies.

(4) Since the half-power beam width (HPBW) of the NRO 45-m telescope
was $15''$, galaxies whose position error listed in the NASA Extragalactic
Database (NED) is less than $10''$ were selected.
We also cross-checked the position using the DSS images.
The 15$''$-beam corresponds to $\sim30 $ kpc at $z \sim 0.1$.
Hence, we may safely assume that most of CO-emitting disks are covered
by the beam.

(5) Since linewidths of the objects were expected to be about 200 to  500 \kms,
we selected galaxies with recession velocity accuracy better than 100 \kms,
in order to fit the band width (250 MHz) corresponding to $650(1+z)$ \kms.
The recession velocity was taken from the IRAS  redshift surveys by
Strauss et al.  (1992) and Fisher et al.  (1995).

The CO and far-infrared properties of individual objects
are described in detail in Tutui et al (2000).
The objects were selected by their isolated and normal morphology.
This makes contrast to the existing CO-observations from the
literature, which are mostly for interacting/merging systems.
Using these new data, Tutui et al (2000) have discussed the molecular-gas and
dust properties of late-type galaxies at intermediate redshift.
We stress that our sample presents the deepest CO observations of
non-intreacting/non-merging IRAS galaxies at intermediate redshift.
The galaxies are shown to have normal star-formation efficiency,
normal color-color diagrams, and not particularly strong nuclear activity.
However, they show smaller gas-to-dust ratio than usual (brighter)
IRAS galaxies.
Nevertheless, our IRAS selected sample could contain starburst galaxies,
which would have higher luminosity comared to normal galaxies.
However, we note that van Driel et al.(1995) have examined the TF relation
for IRAS selected galaxies, and found little difference from the TF relation
of normal galaxies.
Since galaxies rich in CO are often rich in HI, one might wonder
how mamy galaxies in this sample are detected in HI.
However, in so far as we have checked by NED, there have been no
HI detection.

\subsection{CO-Line Observations}

The observataions of the $^{12}$CO($J=1-0$) line were carried out using
the 45-m telescope at the Nobeyama Radio Observatory as a long-term
projects at NRO.
The observations were carried out in January and December in 1994,
January, March and December in 1995, January, February and December in 1996,
and in January in 1997.
The HPBW of the NRO 45-m telescope was
15\hbox{$^{\prime\prime}$} ~at the frequency of $^{12}$CO($J=1-0$) ~line,
and the aperture and main-beam efficiencies were $\eta_{\rm a}$
= 0.35 and $\eta_{\rm mb}$ = 0.50, respectively.
As the receiver frontends, we used cooled SIS
(superconductor-insulator-superconductor) receivers.
The receiver backends were 2048-channel wide-band
acousto-optical spectrometers (AOS) with the band width of
250 MHz, which corresponds to a velocity coverage of
$ 650(1+z) ~{\rm km~s^{-1}}$, or 650 ($z=0$) to 780 \kms\ ($z=0.2$).

The center frequency was set at the 1024-th channel, which
corresponded to $115.271204~(1+z)^{-1}$ for each galaxy.
The system noise temperature was 300 $-$ 800 K in the single side band
at the observing frequencies.
Calibration of the line intensity was made using an absorbing
chopper in front of the receiver, yielding an antenna temperature
($T_{\rm A}^{\ast}$), corrected for both the atmospheric and antenna
ohmic losses.
Intensity scale of $T_{\rm A}^{\ast}$ was converted to the main-beam
brightness temperature by $T_{\rm mb}=T_{\rm A}^{\ast}/ \eta_{\rm mb}$.
Subtraction of  sky emission was performed
by the on-off position switching, and an offset of off-position
was 5$\hbox{$^\prime$}$  far from on-position.
Antenna pointing of the NRO 45-m telescope was done by
observing nearby SiO maser sources at 43 GHz every 60 to 90 minutes,
where the two receiveres (115 and 43 GHz) are well aligned for this perpose.
The pointing accuracy was better than $\pm 4\hbox{$^{\prime\prime}$}$
during the observations.
The total observation time for the on/off position integrations
and pointing was about 2 to 9 hours for individual galaxies, and
the on-source integration time for each galaxy was 1 to 3 hours.

After flagging bad spectra, subtraction of the baseline
was performed by applying the standard procedure of linear-baseline fitting
at both edges of individual spectra.
Adjacent channels were binned to a velocity resolution of
10 $~{\rm km~s^{-1}}$.
The rms noise of the resultant spectra at velocity resolution of 10
$~{\rm km~s^{-1}}$ was 2 $-$ 5 mK in $T_{\rm A}^{\ast}$.

\subsection{CO-line Profiles and Linewidths}

We observed 51 galaxies at intermediate redshift of $cz \sim
10,000 - 50,000 ~{\rm km~s^{-1}}$ and obtained
CO-line profiles of 17 galaxies, as listed in Table 1.
The observed CO-line profiles are shown in Fig. 1.
These galaxies are the deepest CO-line sample of IRAS galaxies with
non-interacting, normal morphology (Tutui et al.  2000).
The detection probability is not high compared to the LIR galaxies,
because the sample consists of very distant galaxies of normal morphology.
CO luminosities for an assumed Hubble constant
of about  60 \kms| Mpc$^{-1}$, as will be obtained in the later
section, are comparable to those of high-mass spirals, and
estimated molecular hydrogen mass of the galaxies is of the order
of $10^9\Msun$ for a conversion factor of
$2\times10^{20}$ H$_2$ [K \kms]$^{-1}$.

--- Table 1 ---

--- Fig. 1 ---

We determined the CO linewidths at the 20\% level of the peak intensity.
Although the line profiles are noisy with typical S/N ratios of about
10, the linewidths are determined within the error of
$\Delta W \sim 10$ to 20 km s$^{-1}$.
Since the expected line shapes from rotating disks are not straightforward,
like gaussian, no automatic algorithm has been applied.
Instead, we judged the linesby eyes, and the errors were taken
to be the uncertainty in the edge channels of lines.
One channel after smoothing, as in the figure, is 10 km/s.
Three galaxies, IRAS 02411+0354, IRAS 07243+1215
and IRAS 14210+4829, are found to be too crude for TF analyses.
The CO linewidth is defined as
\begin{equation}
W_{\rm obs} \equiv c~ \frac{\Delta \nu}{\nu_{\rm obs}},
\end{equation}
where $\Delta \nu$ is the linewidth in observed frequency
and $\nu_{\rm obs}$ is the observed center frequency of
the spectrometer.
The results of the observations are listed in Table 2.

--- Table 2 ---

%%%%%%%%%%%%%%%%%%%%%%%%%%%%%%%%
%  Photometry Observations
%%%%%%%%%%%%%%%%%%%%%%%%%%%%%%%%

\subsection{H and J-band Photometry}

Besides the CO-line observations, we have obtained near-IR (NIR)
photometry in the $J$ and  $H$-bands (Aaronson et al. 1980).
In these NIR bands, the interstellar extinction in our Galaxy
and target galaxies is greatly reduced compared to $B$-band.
This is particularly helpful for the present galaxies,
which were selected from IRAS bright galaxies for CO-line detections.
NIR luminosity of IRAS Minisurvey galaxies is shown to be not
significantly enhanced compared to the $RSA$ sample
(van Driel et al. 1995).

Surface photometry observations  were made
in 1999 January in $J$- (1.25 $\mu\rm m$) and $H$-band (1.65 $\mu\rm m$)
using OASIS, a NIR spectroscopic and imaging camera
attached to the Cassegrain focus of the 1.88-m reflector at the Okayama
Observatory.
The detector was NICMOS-3 which consists of 256 $\times$ 256 pixels.
The seeing size was $1.7'' - 2.1''$ (FWHM), and the sky level was
typically 16 mag in $J$-band and 14 mag in $H$-band.
One pixel corresponded to $0.97\hbox{$^{\prime\prime}$}$ ,
and the field-of-view was approximately $4\hbox{$^\prime$}$.
The exposure time per one frame was 30--80 seconds for $J$-band
and 10--30 seconds for $H$-band, and total integration time was about
20 minutes for each band per one galaxy.
The flux calibration was performed using the standard stars
in $J$ and $H$-bands presented by Hunt et al.  (1998)
before and after each observation of galaxies.
Standard data processing (dark current subtracting, image shifting/combining
and flat fielding) was performed with the IRAF software package,
and subsequent surface photometry and image processing
(e.g. sky subtraction and flux measurement) was performed with the 
SPIRAL package developed at the Kiso Observatory and
incorporated into the IRAF system (Hamabe \& Ichikawa 1992).
Figure 2 shows the obtained J-band images.

--- Fig. 2 ---

We obtained $J$ and $H$-band photometric observations for
12 galaxies out of the 17 CO-detected galaxies,
and measured the total magnitude.
However, IRAS 17517+6422 was found to be an interacting galaxy,
we did not include this galaxy in our TF analysis.
Also, PG0157+001 was found to be a quasar, and was not included.
Hence, the TF analysis was applied to the rest 10 galaxies, as listed
in Table 4.

We measured the integrated magnitude of galaxies within a circular annuli
outward from the galaxy center with 1 pixel interval. We define the total
magnitude as the converged value of the growth curve.
The total magnitude thus defined have the error as large as 0.1 mag since
the galaxy morphology is not well determined (de Vaucouleurs 1991).

The following corrections were applied to the observed total magnitude $m_T$
to obtain corrected total magnitude $m_T^0$ is written as
\begin{equation}
m_T^0 = m_T - A_i - A_G - K(z),
\label{eq_mT0}.
\end{equation}
Here, $A_i$ is the internal extinction within the target galaxy, and was
calculated from B-band value taken
from RC3 (de Vaucouleurs et al. 1991; Watanabe et al. 1998)
using a formula
$A_i^J = 0.21 A_i^B$ and  $A_i^H = 0.13 A_i^B$
(Rieke \& Lebofsky 1985).
The Galactic extinction $A_G$ is taken from Berstein \& Heiles (1982).
The $K$-correction, a correction for a redshifted wavelength of
observed passbands, is estimated by assuming the spectral model
of Poggianti (1997) for an Sc galaxy.
The corrected total magnitude $m_T^0$ in Eq.(2)
was further corrected for the evolution,
$E$-correction, adopted from  Poggianti (1997).
We defined the total magnitude corrected for $K$ and $E$-corrections
as $m_T^1$,
\begin{equation}
m_T^1 = m_T - A_i - A_G - K(z) -E(z).
\end{equation}
The errors of the corrected magnitudes are assumed
to be the same as those of the observed value $m_T$, 0.1 magnitude.
The value of these corrections and the corrected magnitude
are also listed in Table 3.
The errors of the thus calculated corrected magnitudes are assumed
to be the same as those for the observed value $m_T$,
which is 0.1 magnitude.

--- Table 3 ---

%%%%%%%%%%%%%%%%%%%%%%%%%%%%%%%%%%%
%  Hubble constant
%%%%%%%%%%%%%%%%%%%%%%%%%%%%%%%%%%%

\section{Determination of the Hubble constant}

\subsection{Inclination correction}

Since B-band images would be affected by recent star formation
which disturbs the isophote in bluer bands,
 we used R-band images taken from the STScI Digitized Sky Survey
 for determination of the inclination angle, using the
  conventional formula given by Hubble (1926):
\begin{equation}
\cos^2 i = \frac{(b/a)^2 - {\sf q}_{0}^2}{1-{\sf q}_{0}^2}
\end{equation}
\noindent
where $b/a$ is the minor-to-major axial ratio
and ${\sf q}_0$ is an intrinsic axial ratio fixed to 0.2.
The disk images of the sample galaxies in $J$ and $H$ bands are too
faint against the high sky background to fit the major and minor axes.
Therefore, we adopt the deeper optical $R$ band images for obtaining the
inclination.

\subsection{Conversion of CO to HI linewidths}

The inclination-corrected CO-line velocity width is then
calculated as
\begin{equation}
W^c_{\rm CO} = \frac{W_{\rm obs}}{\sin i},
\label{eq_Wc}
\end{equation}
where $i$ is the inclination angle,
and the suffix $c$ denotes the corrected linewidth.
Tutui \& Sofue (1999) and Tutui (1999) showed,
that the CO linewidth is not entirely identical with the HI linewidth:
\begin{equation}
W^c_{\rm HI} = 0.76 ~ W^c_{\rm CO} + 83.8.
\label{eq_HI2COLW_ch.TFR}
\end{equation}
The HI linewidth is slightly larger than CO width for slowly rotating galaxies,
and vice versa for fast rotating galaxies.
This is caused by different gas distribution in a galaxy and
by velocity dependence on a shape of the rotation curve.
These corrected linewidths are listed in Table 4.
Note that, according to this equation,
$W^c_{\rm HI}$ and $W^c_{\rm CO}$ coincide within the
error for galaxies with typical line widths of around 350 \kms.

--- Table 4 ---

\subsection{Adopted Tully-Fisher relations}

We then apply the HI Tully-Fisher relation in $J$ and $H$-bands
derived by Watanabe et al (2001) for the thus corrected values of $W^c_{\rm HI}$.
Watanabe et al. (2001) calibrated the TF relations using the local calibrators
with the Cepheid distances observed by HST,  where the zero-point calibrations
were performed for the calibrators observed with the Kiso Observatory
105-cm Schmidt telescope.
We obtain the total matnitude, the inclination, the extinction correction,
and the line with in the same way as those of the calibrators.
\begin{equation}
 M_{J_T} = -8.48 (\pm 0.85)(\log W_{\rm HI}^{c} -2.5) -22.13 (\pm 0.39) 
\end{equation}
\begin{equation}
 (\sigma=0.30),
%M_{J_{\rm T}} = (-8.59 \pm 0.99)(\log W_{\rm HI}^{c} -2.5) -21.96 \pm 0.17,
\label{JbandTFW99}
\end{equation}
and
\begin{equation}
M_{H_T} = -7.54 (\pm 0.76)(\log W_{\rm HI}^{c} -2.5) -22.95 (\pm 0.35) 
\end{equation}
\begin{equation}
(\sigma=0.28).
%M_{H_{\rm T}} = (-8.34 \pm 0.77)(\log W_{\rm HI}^{c} -2.5) -22.64 \pm 0.13.
\label{HbandTFW99}
\end{equation}
%for $log W_{HI}^c > 2.45$.

\subsection{Determination of Hubble constant}

Luminosity distance derived from the TFR is given by the distance
modulus, $m - M$, as
\begin{equation}
\log D_L = -5 + \frac{1}{5} (m - M)~~ {\rm (Mpc)},
\end{equation}
where $m$ is apparent total magnitude.
The luminosity distance is related to the Hubble constant as
\begin{equation}
D_L = \frac{c}{H_0 q_{\rm 0}^2} \left\{ q_{\rm 0} z + (q_{\rm 0}-1)(\sqrt{2 q_{\rm 0} z + 1} -1)
\right \},
\label{DL}
\end{equation}
where $q_{\rm 0}$ is the deceleration parameter, and we take $q_{\rm 0}=0.5$.
Then, the Hubble ratio  is written as,
\begin{equation}
H_0 = \frac{2c}{D_L} (z+1 - \sqrt{z+1}),
\end{equation}
For small redshift, $z \ll 1$,  this relation reduces, of course, to
$H_0 = cz / D_L$.
The obtained Hubble ratios for all sample galaxies is listed in
Table 5.

--- Table 5 ---

The distances and Hubble ratios derived from the Tully-Fisher
relations are listed in Table 5, dand the J and H-band results after
$K$ correciton are plotted in Fig. 3.
We, then, calculated the mean Hubble constant in the observed redshift 
range by weight-averaging the individual values, which the results
are listed in Table 6.
Thereby, values exceeding 3$\sigma$ dispersion of the plots are excluded.
Namely, two galaxies, IZw 23, CGCG 1113.7+2936, as well as the quasar PG 0157
were excluded from the sample.
Note that the plots in Fig. 3 and  are more scattered
for lower redshift galaxies, for which the beam size (15$''$)
could not cover the whole CO emitting regions.

---- Table 6 ----

---- Fig. 3 ----

The weighted mean value of the Hubble ratios from the $J$-bands results
after $K$ correction is obtained to be 
$H_0 = 60 \pm 10 ~{\rm~km~s{}^{-1}~Mpc{}^{-1}}$.
The $H$ band result leads to $H_0 = 53 \pm 13 ~{\rm~km~s{}^{-1}~Mpc{}^{-1}}$.
The Hubble ratios for $J$ and $H$-band are almost equal
for some galaxies, confirming that the uncertainty in surface photometry
in $J$ and $H$-band is small.
The $K$-and-$E$ corrected Hubble constant is determined as
$H_0 = 58 \pm 10 ~{\rm~km~s{}^{-1}~Mpc{}^{-1}}$ in $J$-band,
and $H_0 = 52 \pm 12 ~{\rm~km~s{}^{-1}~Mpc{}^{-1}}$ in $H$-band,
which are only 3\% smaller than the value after the $K$ correction alone. 
 
\subsection{Errors}

Given a TF relation,
the error in the absolute magnitude $\Delta M$ arises from the errors
in the CO linewidth $\Delta W_{\rm obs}$, inclination $\Delta i$,
and redshift $\Delta z$, which propagate to each other as follows.
\begin{equation}
\Delta M = \sqrt{
\left( \frac{\Delta W_{\rm obs}}{W_{\rm obs}} \right)^2
+ \left( \frac{\Delta i}{{\rm tan} i} \right )^2
+ \left( \frac{\Delta z}{1+z} \right)^2
} \cdot |A| \log e,
\end{equation}
where $A$ is the slope of the TFR.
The error in the luminosity distance is, then,
related to the error in the absolute magnitude as above,
and the error in the apparent magnitude measurement:
\begin{equation}
\frac{\Delta D_L}{D_L} = \frac{\Delta (m-M)}{5 \log e}
= \frac{1}{5 \log e} \sqrt{\Delta m^2 + \Delta M^2}
\end{equation}
and
\begin{equation}
\frac{\Delta D_L}{D_L} = \frac{1}{5 \log e} \times
\end{equation}
\begin{equation}
\sqrt{
\Delta m^2 + \left\{
\left( \frac{\Delta W_{\rm obs}}{W_{\rm obs}} \right)^2
+ \left( \frac{\Delta i}{{\rm tan} i} \right )^2
+ \left( \frac{\Delta z}{1+z} \right)^2
\right\} |A|^2 (\log e)^2.
}
\end{equation}
For small redshift,
the error in Hubble ratio is written as,
\begin{equation}
\frac{\Delta H_0}{H_0} = \sqrt{
\left( \frac{\Delta z}{z} \right)^2
+ \left( \frac{\Delta D_L}{D_L} \right)^2.
}
\end{equation}

The contribution of the errors in total magnitude
$\Delta m$ and redshift $\Delta z$ to the error of the
distance estimation is much smaller
than that of the errors of linewidth $\Delta W_{\rm obs}$ and
inclination $\Delta i$, which was measured to be about $4^\circ.8$ 
for all galaxies.
Therefore, the first and 4th terms in the square root of Eq. (14) are
negligible compared to the other terms.
The error in the resultant Hubble ratio increases with decreasing linewidth.
The beam-size effect is crucial for lower redshift galaxies:
It may happen that the telescope beam cannot cover the entire CO extent
for large angular diameter galaxies.
The telescope beam (15$''$) corresponds to 36 kpc for
a galaxy at $cz=30,000$ \kms for our resulting Hubble constant of about
60 \kms\ Mpc$^{-1}$. This diameter covers the substantial portion of the
interstellar gas disks, and the detected CO line well represents the
maximum velocity width.
However, for lower redshift galaxies than 15,000 \kms, for example,
the beam might not cover the entire disk, which may result in
underestimation of the line width, and therefore, underestimation
of the Hubble ratio.
In fact, the scatter of the plot in Fig. 3 increases with decreasing redshift,
implying the beam effect for low redshift galaxies would not be negligible.
The result would be, therefore, more reliable for galaxies at $cz>20,000$ \kms
in the present case for the 45-m telescope.

%%%%%%%%%%%%%%%%%%%%%%%%%%%%%%%
%  Discussion
%%%%%%%%%%%%%%%%%%%%%%%%%%%%%%%

\section{Discussion}

Peculiar velocities of galaxies, either individual or due to
large-scale structures in clusters and networks, contribute
significantly to the scatter in Hubble ratios.
Therefore, in order to obtaine a more accurate Hubble constant,
it is important to go to higher redshifts where the
peculiar velocities are small compared to the recession velocity.
On the other hand, the evolution effect becomes significant at higher
redshift, which is still not easy to correct for by the current models.
These confilicting requirements can be optimized by applying the TFR to
galaxies at intermediate redfhits.
For such purposes, the CO TFR would potentially be an alternative to
HI TFR, particularly when larger and more sensitive facilities such
as the ALMA becomes available in the future.

We have thus shown that the CO TFR is quite possible in the cosmological
distances at $cz \sim 10,000 - 50,000 ~{\rm km~s^{-1}}$,
although the detectability and the accuracy are not satisfactory yet.
Nevertheless, we were able to determine the mean Hubble constant in the
space between $cz \sim 10,000$ and $35,000 ~{\rm km~s^{-1}}$ to be
$H_0=60\pm10 {\rm km~s^{-1} Mpc^{-1}}$ from the $J$-band TFR.

Our value is consistent within the error with the recent value
$H_0=65$ \kmsmpc\ from HI TFR at $cz<12000$ km s$^{-1}$
(Watanabe et al. 1998), and is slightly smaller thatn 
that from the HST key project, $ H_0=71$ \kmsmpc, for galaxies within 
25Mpc and clusters within 10,000 km s$^{-1}$  (Mould et al. 2000).
Our value is also consistent with those obtained at
similar redshifts using Type I supernovae
(Sakai et al. 2000; Branch et al. 1998), those
using the Sunyaev-Zel'dovic (1972) effect (Hughes and Birkinshaw  1998),
and those from gravitationally lensed QSOs (Williams et al. 2000).
In Fig. 4 we plot the recent values of Hubble constants derived from
various methods, following Okamura (1999).
Vertical bars indicate errors, and horizontal bars the
redshift coverage.

--- Fig. 4 ---

We emphasize that the present CO-line method provides
us with an alternative, new tool to estimate the distances of
intermediate-redshift galaxies
in the scheme of the established Tully-Fisher relation.
The thus obtained Hubble constant can be directly compared with
those obtained for nearer galaxies using the TFR.

We finally stress that the CO-NIR TFR will be a promising tool for
high-redshift galaxies, which are supposed to be dusty,
if the evolutionry effect can be properly corrected.
The CO-NIR TFR as proposed in this paper will become one of the 
methods for cosmological distance estimates at high redshifts using 
the coming largest mm- and sub-mm wave facility, ALMA, by which higher 
transition CO lines will be observed with much higher sensitivity.
The present project using the 45-m telescope
has been performed in order to establish a  methodology
of the CO TFR, and to evaluate the feasibility for future projects
using the Nobeyama mm-wave Array and ALMA.

%%% Acknowledgments
\vspace{1pc}\par
YT and MH  would like to thank the Japan Society for
the Promotion of Science for the financial support
for young researchers.
IRAF is distributed by the National Optical Astronomy Observatories, which
is operated by the Association of Universities for Research in Astronomy, Inc.
(AURA) under cooperative agreement with the National Science Foundation.

\section*{References}

\re Aaronson, M., Huchra, J., Mould, J.\ 1980, ApJ, 237, 655

\re Berstein, D., Heiles, C.\ 1982, AJ,  87, 1165

\re Branch, D. 1998 ARA\&A 36, 17.

\re Casoli, F., Boisse, P., Combes, F., Dupraz, C.\ 1991, A\&A, 249, 359

\re de Vaucouleurs, G., de Vaucouleurs, A., Corwin, H.G.,Jr., Buta, R.J.,
	Paturel, G., Fouqu\'e, P.\ 1991
	{\it Third Reference Catalog of Bright Galaxies (RC3)}
	(New York: Springer-Verlag)

%\re 	de Vaucouleurs,	G., Buta, R.\ 1983, AJ, 88, 939

\re Dickey, J. M., Kazes, I.\ 1992, ApJ 393, 530

\re Eastman, R. G., Schmidt, B. P., Kirshner, R.\  1996, ApJ, 466, 911

\re Fisher, K. B., Huchra, J. P., Strauss, M. A.,
Davis, M., et al.\ 1995, ApJS, 100, 69

\re Furuzawa, T., Tawara, Y., Yamashita, K., Miyoshi, S.\ 1997,
	IAU Symp. 183, 70

\re Giovanelli, R., Haynes, M. P.\ 1985, ApJ, 292, 404

\re Giovanelli, R., Haynes, M. P., Salzer, J. J.,
Wegner, G., Da Costa, L. N., Freudling, W.\  1995, AJ, 110, 1059

%\re Giovanelli, et al. 1997 AJ 113, 53

\re Hamabe, M., and Ichikawa, S. 1992, in A.S.P. Conference Ser. Vol. 25,
Proc. Astronomical Data Analysis Software and Systems I,
ed. D. M. Worrall et al. (San Francisco: Astronomical Society of the Pacific),
325

\re Haynes, M. P., Giovanelli, R.\ 1986, ApJ, 306, 466

\re Hubble, E. P.\ 1926, ApJ, 64, 321

\re Hughes, J. P., and Birkinshaw, M. 1998 ApJ 501,  1

\re Hunt, L. K., Mannucci, F., Testi, L., Migliorini, S.,
Stanga, R. M., Baffa, C., Lisi, F., Vanzi, L.  1998 AJ, 115, 2594.

\re Jacoby, G. H.\ 1997,
	in {\it The Extragalactic Distance Scale}, eds. Livio, M. et al.
	(Cambridge: Cambridge University Press) p.197

\re Kenney, J. D., Young, J. S.\ 1988, ApJS 66, 261

%\re Kundic, T., Turner, E. L., Colley, W. N., Gott, J. R., III,
%Rhoads, J. E., 	Wang, Y. et al.\ 1997, ApJ, 482, 75

\re Lavezzi, T. E. and Dicky, J. 1998 AJ 116, 2672

%\re Malmquist, K.G.\ 1920, in {\it A Study of the stars of spectral type A.}(Lund: Scientia)

\re Mould, J. R., Huchra, J. P.,
	Freedman, W. L., Kennicutt, R. C. Jr.,
	Ferrarese, L., Ford, H. C., et al. 2000, ApJ, 529, 786

\re Okamura, S.\ 1999,
	in {\it Milky Way Galaxy and the Universe of Galaxies}
	(Tokyo: University of Tokyo Press) p.203

\re Poggianti, B. M.\ 1997, A\&AS, 122, 399

\re Rieke, G. H., Lebofsky, M. J.\ 1985, ApJ, 288, 618

\re Riess, A. G., Press, W. H., Kirshner, R. P.\ 1995, ApJ, 438, L17

\re Sakai, S., Mould, J. R., Hughes, S. M. G., Huchra, J. P.,
Macri, L. M., Kennicutt, R. C. Jr. et al.\ 2000, ApJ, 529, 698

\re Sandage, A., Saha, A., Tammann, G. A., Labhardt, L.,
Panagia, N., Macchetto, F. D.\ 1996, ApJ, 460, L15

\re Sch\"oniger, F., Sofue, Y., 1994, A\&A 283, 21

\re Sch\"oniger, F., Sofue, Y., 1997, A\&A 323, 14

\re Sofue, Y., 1992, PASJ 44, L231

\re Sofue, Y., Sch\"oniger, F.,  Honma, M.,  Tutui, Y.,  Ichikawa,, T.,
        Wakamatsu, K., Kazes, I. \& Dickey, J.
        1996 PASJ 48, 657-670

\re Strauss, M. A., Huchra, J. P., Davis, M., Yahil, A. et al.
\ 1992, ApJS, 83, 29

\re Sunyaev, R.A. \& Zel'dovich, Ya.B. 1972,
	in {\it Comm. Astrophys. Space. Phys.,} 40, 173

\re Tanvir, N. R., Shanks, T., Ferguson, H. C., Robinson, D. R. T.\ 1995,
Natur, 377, 27

\re Tonry, J. L.\ 1991, ApJ, 373, L1

\re Tutui, Y. \&  Sofue, Y.\ 1997, 	A\&A 326, 915

\re Tutui, Y., \&  Sofue, Y.\ 1999, A\&A, 351, 467

\re Tutui, Y.\ 1999, in {\it the Ph.D. thesis of the University of Tokyo}

\re Tutui, Y., Sofue, Y., Honma, M.,  Ichikawa, T., Wakamatsu, K.
2000 PASJ 52, 803.

\re Van Driel, W., Van Den Broek, A. C., Baan, W.\ 1995, ApJ, 444, 80

\re Watanabe, M., Ichikawa, T., Okamura, S.\ 1998, ApJ, 503, 553

%\re Watanabe, M., Yasuda,N., Itoh, N.\ 1999, private communication

\re Watanabe, M., Tasuda, N., Itoh, N. Ichikawa, T., \& Yanagisawa, K.
2001, ApJ, in press

\re Williams L. L. R. and Saha, P. 2000, AJ, 119, 439

\newpage

Figure Captions

Fig. 1. CO line profles for 17 galaxies. The horizontal bars
indicate the measured line widths. Galaxiese with asterisques
are for marginal and non-detection, which are not included in
the distance estimation.

Fig. 2. J and H-band  images of the measured galaxies.
mage sizes are $1'\times 1'$ except for for NGC 6007 of $2'\times 2'$.
Faintest isophote is 22 mag arcsec$^{\rm -2}$ for $J$-band, and 20
mag  arcsec$^{\rm -2}$ for $H$-band.
Contour intervals is 1 mag arcsec$^{\rm -2}$. Top is to the north,
and left to the east. 

Fig. 3. Hubble ratios plotted against redshifts for $J$-
(filled circles) and $H$-band (diamonds), after $K$ correction.  
Note the smaller scatter for  larger redshifts than $\sim 20000$ \kms.

Fig. 4. Hubble constants determiend by various methods (Okamura 1999).
The references are: SBF (surface brightness fluction: Tonry 1991);
PNLF (planetary nebulae luminosity function: Jacoby 1997);
HST (The HST key project: Sakai et al. 2000);
Cepheid (Tanvir et al. 1995);
TF (Watanabe et al. 1999);
SN II (Type II SN: Eastman et a. 1996);
SN Ia(R) (Type Ia SN: Riess et al. 1995);
SN Ia(S) (Type Ia SN: Sandage et al. 1996;
GL (Gravitaional Lens: kundic et al. 1997);
SZ (Sunyaev-Zel\'dovich effect: Furuzawa 1997); and
COTF (This work).

\newpage

\def\arcsec{''}
\def\arcmin{'}
\begin{table*}[t]
% \small
%%%%%%%%%%%%%% Table 1 %%%%%%%%%%%%%%%
\begin{center}
Table~1.\hspace{4pt} Galaxies detected in the CO-line
using the NRO 45-m telescope.\\
\end{center}
\vspace{6pt}
\begin{tabular*}{\textwidth}{@{\hspace{\tabcolsep}
\extracolsep{\fill}}lccccccl}
\hline\hline
%%%%%
Galaxy (IRAS ID)& RA$_{1950}$	& Dec$_{1950}$	& $cz$	&$z$ & $D_{cz}$ & J-band size\\
        & ~h~~m~~s	& $~~{}^{\circ}~~\arcmin~~\arcsec$ & km s$^{-1}$ & & Mpc& $''\times''$\\
\hline

PG 0157+001 (01572+0009)	& 01 57 16.3	& +00 09 09	& 48869 & 0.16301 & 677	&24x15\\
IRAS 02185+0642  		& 02 18 40.3	& +06 43 03	& 29347 & 0.09789 & 401	&19x17 \\
IRAS 02411+0354$^{\dagger}$ 	& 02 41 09.3	& +03 53 56	& 43050 & 0.14360 & 594	& \\
IRAS 07243+1215$^{\dagger}$     & 07 24 20.6	& +12 15 09	& 28204 & 0.09408 & 385	& \\
I Zw23 (09559+5229)	& 09 56 01.0	& +52 29 48	& 12224 & 0.04077 & 165	& 31x25\\
CGCG 1113.7+2936  (11137+2935)	& 11 13 47.1	& +29 35 58	& 13880 & 0.04630 & 187 & 50x31\\
IC 2846	(11254+1126)	& 11 25 24.8	& +11 26 01	& 12294 & 0.04101 & 166	& 46x27\\ %Sp;LIN HII
IRAS 14060+2919	 		& 14 06 04.9	& +29 18 59	& 35060 & 0.11695 & 481	& 17x11\\
CGCG 1417.2+4759 (14172+4758)& 14 17 14.8& +47 59 00	& 21465 & 0.07160 & 291	& 25x21\\ % SBcd\\
IRAS 14210+4829$^{\dagger}$  	& 14 21 06.2	& +48 29 59	& 22690 & 0.07569 & 308	& \\ % Sp
CGCG 1448.9+1654 (14488+1654)	& 14 48 54.5	& +16 54 02	& 13700 & 0.04570 & 185	& 23x19\\ % Sab
NGC 6007 (15510+1206)	& 15 51 01.6    & +12 06 27	& 10547 & 0.03518 & 142	& 85x43\\
IRAS 16533+6216	               & 16 53 19.8	& +62 16 36	& 31808 & 0.10610 & 435	& 13x12\\ % E
PGC 60451 (17300+2009)	& 17 30 00.6	& +20 09 49	& 14989 & 0.05000 & 202	& \\ % Sc
IRAS 17517+6422	 	        & 17 51 45.0	& +64 22 14	& 26151 & 0.08723 & 356	& \\ % E
IRAS 23389+0300	 		& 23 38 56.9	& +03 00 48	& 43470 & 0.14500 & 600	& \\
IRAS 23420+2227                & 23 42 00.6     & +22 27 50     & 26022 & 0.08680 & 354	& 15x10\\[4pt]
\hline
\end{tabular*}
\vspace{6pt}\par\noindent
Col.(1): Galaxy name. A dagger denotes a galaxy
of marginal detection. Col.(2): Alias of the galaxy name as the IRAS
catalog name.
Col.(2) and (2): Coordinates in B1950.
Col.(4): Heliocentric velocity.
Col.(5): Heliocentric redshift from Fisher et al. (1995).
Col.(6): Distance derived from the redshift,
assuming $H_0 = 75 ~{\rm~km~s{}^{-1}~Mpc{}^{-1}}$ and $q_0 = 0.5$.
Col.(7): J-band angular size in arc seconds for galaxies observed in the
present photometry imaging.
\end{table*}

\newpage
%%%%%%%%%%%%%%% Table 2 %%%%%%%%%%%%%%%%
\begin{table*}[t]
\small
\begin{center}
Table~ 2. CO-line observation results
using the NRO 45-m telescope.
\end{center}
\vspace{6pt}
\begin{tabular*}{\textwidth}{@{\hspace{\tabcolsep}
\extracolsep{\fill}}lccccrrr}
\hline\hline
%%%%
  Galaxy	& $cz$	   & $t_{\rm int}$ & r.m.s.& $W_{\rm obs}$ &
$T_{\rm A}^{\ast}$ & ${I}_{\small{\rm CO}}$ & \\
  		& $~{\rm km~s^{-1}}$ & min. & mK & $~{\rm km~s^{-1}}$ & mK & $~{\rm K~km~s{}^{-1}}$ &
\\
\hline
%%%%
PG 0157+001 &    48869 &   90 & 2 		& 300 $\pm$ 15 &     15 &  5.11 &  (0.22) \\
IRAS02185+0642 &    29347 &   90 & 4 		& 356 $\pm$ 15 &     40 & 13.05 &  (0.48) \\
IRAS02411+0354$^{\dagger}$ &    43050 &   60 & 5 & ----      &      8 &   1.90 &  (0.66) \\
IRAS07243+1215$^{\dagger}$ &    28204 &   90 & 3 & ----      &      6 &   1.79 &  (0.28) \\
IZw23 &    12224 &   60 & 4 		& 104 $\pm$ 5  &     39 &  4.82 &  (0.26) \\
CGCG1113.7+2936 &    13880 &   60 & 5 		& 248 $\pm$ 20 &     14 &  4.33 &  (0.55) \\
IC2846 &    12294 &   30 & 4 		& 360 $\pm$ 20 &     19 &  4.78 &  (0.48) \\
IRAS14060+2919 &    35060 &  120 & 3 		& 376 $\pm$ 20 &      8 &  1.91 &  (0.37) \\
CGCG1417.2+4759 &    21465 &  180 & 3 		& 293 $\pm$ 15 &     13 &  4.66 &  (0.32) \\
IRAS14210+4829$^{\dag}$ &    22690 &   60 & 4 & ----     &     11 &  3.08 &  (0.54) \\
CGCG1448.9+1654 &    13700 &   60 & 4 		& 282 $\pm$ 10 &     28 &  7.03 &  (0.42) \\
NGC6007 &    10547 &   60 & 7 		& 347 $\pm$ 10 &     28 & 11.24 &  (0.82) \\
IRAS16533+6216 &    31808 &   90 & 3 		& 217 $\pm$ 10 &     10 &  3.54 &  (0.28) \\
PGC60451 &    14989 &   60 & 7 		& 459 $\pm$ 15 &     20 & 10.35 &  (0.95) \\
IRAS17517+6422 &    26151 &   90 & 5 		& 480 $\pm$ 20 &     19 &  4.00 &  (0.69) \\
IRAS23389+0300 &    43470 &  180 & 3 		& 272 $\pm$ 20 &      8 &  3.92 &  (0.31) \\
IRAS23420+2227 &    26022 &   60 & 7 		& 308 $\pm$ 20 &     15 &  4.80 &  (0.49) \\[4pt]
\hline
\end{tabular*}
\\
\vspace{6pt}\par\noindent
Col.(1): Galaxy name. A dagger denotes a galaxy of
marginal detection. Col.(2): Redshift in $cz$.
Col.(3): Integration time of on-source in minute.
Col.(4): Root mean square of antenna temperature after binning of
10 $~{\rm km~s^{-1}}$ in emission-free region of the spectrum.
Col.(5): Observed CO linewidth defined as the full width at 20\% of the maximum intensity.
Col.(6): Antenna temperature at the peak level intensity.
Col.(7): Integrated intensity corrected for the main beam efficiency.
Col.(8): Uncertainty of 1 $\sigma$ in the integrated intensity.
\end{table*}

\newpage
%%%%%%%%%%%%%%%%% Table 3 %%%%%%%%%%%%%%%%
\begin{table*}[t]
\small
\begin{center}
Table~3.\hspace{4pt} Total magnitude and correction for magnitude.
\end{center}
\vspace{6pt}
\begin{tabular*}{\textwidth}{@{\hspace{\tabcolsep}
\extracolsep{\fill}}lccccccc}
\hline\hline
  Galaxy	& $m_T$ & $A_i$ & $A_G$ & $K$ & $E$ & $m_T^0$ &
$m_T^1$ \\
  		& mag & mag & mag & mag & mag & mag & mag\\
\hline
($J$-band)\\
PG 0157+001      & 14.23  & 0.02 & 0.04	& -0.09 & -0.17 &14.26 & 14.43 \\
IRAS 02185+0642  & 14.37  & 0.02 & 0.04	& -0.06 & -0.10 &14.37 & 14.47 \\
I Zw23           & 12.77  & 0.02 & 0.04	& -0.03 & -0.05 &12.74 & 12.79 \\
CGCG 1113.7+2936 & 13.41  & 0.05 & 0.03	& -0.04 & -0.05 &13.37 & 13.42 \\
IC 2846          & 12.26  & 0.03 & 0.04	& -0.03 & -0.05 &12.21 & 12.26 \\
IRAS 14060+2919  & 15.10  & 0.03 & 0.03	& -0.07 & -0.12 &15.11 & 15.23 \\
CGCG 1417.2+4759 & 13.89  & 0.02 & 0.04	& -0.05 & -0.08 &13.88 & 13.96 \\
CGCG 1448.9+1654 & 14.12  & 0.02 & 0.04	& -0.03 & -0.05 &14.09 & 14.15 \\
NGC 6007         & 12.66  & 0.03 & 0.05	& -0.03 & -0.05 &12.61 & 12.66 \\
IRAS 16533+6216  & 15.16  & 0.02 & 0.05	& -0.07 & -0.14 &15.15 & 15.29 \\
IRAS 23420+2227  & 15.64  & 0.03 & 0.04	& -0.06 & -0.10 &15.63 & 15.73 \\
($H$-band)\\
PG 0157+001      & 13.32 & 0.01 & 0.02 &-0.08 &-0.16 &13.37 & 13.53\\
IRAS 02185+0642  & 13.18 & 0.02 & 0.03 &-0.04 &-0.10 &13.17 & 13.27\\
I Zw23           & 12.26 & 0.01 & 0.03 &-0.02 &-0.04 &12.24 & 12.28\\
CGCG 1113.7+2936 & 12.34 & 0.03 & 0.02 &-0.02 &-0.05 &12.31 & 12.36\\
IC 2846          & 11.49 & 0.02 & 0.02 &-0.02 &-0.04 &11.47 & 11.51\\
IRAS 14060+2919  & 14.57 & 0.02 & 0.02 &-0.05 &-0.12 &14.58 & 14.70\\
CGCG 1417.2+4759 & 13.32 & 0.01 & 0.02 &-0.03 &-0.08 &13.31 & 13.39\\
CGCG 1448.9+1654 & 13.98 & 0.02 & 0.02 &-0.02 &-0.05 &13.96 & 14.01\\
NGC 6007         & 12.37 & 0.02 & 0.03 &-0.01 &-0.04 &12.33 & 12.37\\
IRAS 16533+6216  & 14.52 & 0.01 & 0.03 &-0.04 &-0.11 &14.52 & 14.63\\
IRAS 23420+2227  & 14.91 & 0.02 & 0.02 &-0.04 &-0.09 &14.90 & 14.99\\
\hline
\end{tabular*}
\vskip 2mm
\noindent Col.(1): Galaxy name.
Col.(2): Observed total magnitude.
Col.(3): Internal extinction.
Col.(4): The Galactic extinction.
Col.(5): $K$-correction.
Col.(6): $E$-correction.
Col.(7): Corrected total magnitude without $E$-correction
written by $m_T^0 \equiv m_T - A_i - A_G - K$.
Col.(8): Corrected total magnitude
written by $m_T^1 \equiv m_T - A_i - A_G - K - E$.
\end{table*}

\newpage
%%%%%%%%%%%%%%%%%%%%%% Table 4 %%%%%%%%%%%%%%%%
\begin{table*}[t]
\small
\begin{center}
Table~4.\hspace{4pt} Linewidths and total magnitudes.
\end{center}
\vspace{6pt}
\begin{tabular*}{\textwidth}{@{\hspace{\tabcolsep}
\extracolsep{\fill}}lccccccccc}
\hline\hline
Galaxy		& $cz$ & $i$ & $W_{\rm CO}$ & $W^c_{\rm CO}$ &
$W^c_{\rm HI}$ & $m^0_{\rm T} (J)$ & $m^1_{\rm T} (J)$ & $m^0_{\rm T} (H)$ & $m^1_{\rm T} (H)$ \\
		& ${\rm km~s^{-1}}$ & deg & ${\rm km~s^{-1}}$ & ${\rm
km~s^{-1}}$ & ${\rm km~s^{-1}}$ & mag & mag & mag & mag \\
\hline
   IRAS 02185+0642 &  29347 &  33 &  356 &  654 &  581 &  13.85 &  13.95 &  13.17 &  13.27 \\
     I Zw23 &  12224 &  49 &  104 &  138 &  189 &  12.74 &  12.79 &  12.24 &  12.28 \\
    CGCG 1113.7+2936 &  13880 &  58 &  248 &  293 &  306 &  13.37 &  13.42 &  12.31 &  12.36 \\
    IC 2846 &  12294 &  34 &  360 &  644 &  573 &  12.21 &  12.26 &  11.47 &  11.51 \\
   IRAS 14060+2919 &  35060 &  47 &  376 &  514 &  475 &  15.11 &  15.23  &  14.58 &  14.70 \\
    CGCG 1417.2+4759 &  21465 &  44 &  293 &  422 &  404 &  13.88 &  13.96 &  13.21 &  13.29 \\
    CGCG 1448.9+1654 &  13700 &  40 &  282 &  439 &  417 &  14.09 &  14.15  &  13.96 &  14.01 \\
     NGC 6007 &  10547 &  43 &  347 &  509 &  471 &  12.61 &  12.66  &  12.33 &  12.37 \\
   IRAS 16533+6216 &  31808 &  27 &  217 &  478 &  447 &  15.15 &  15.29  &  14.52 &  14.63 \\
   IRAS 23420+2227 &  26022 &  51 &  308 &  396 &  385 &  15.63 &  15.73  &  14.90 &  14.99 \\[4pt]
\hline
\end{tabular*}
\vskip 2mm
\noindent Col.(1): Galaxy name.
Col.(2): Redshift in $cz$.
Col.(3): Inclination. The inclination errors were measured to be about
$4^\circ.8$ for all galaxies, and we adopted the same value through this paper.
Col.(4): CO linewidth.
Col.(5): CO linewidth corrected for the inclination.
Col.(6): Converted CO linewidth corresponding to HI linewidth.
Col.(7): Total magnitude defined as $m_T^0 \equiv m_T - A_i - A_G - K$ for J band.
Col.(8): Total magnitude defined as $m_T^1 \equiv m_T - A_i - A_G - K - E$ for J band.
Col.(9):  $m_T^0 $ for H band.
Col.(8): $m_T^1 $ for H band.
\end{table*}

\newpage
%%%%%%%%%%%%%%%%%%%%% Table 5 %%%%%%%%%%%%%%%%%%
\begin{table*}[t]
\small
\begin{center}
Table~5.\hspace{4pt}  Tully-Fisher distance and Hubble ratio.
\end{center}
\vspace{6pt}
\begin{tabular*}{\textwidth}{@{\hspace{\tabcolsep}
\extracolsep{\fill}}lccccc}
\hline\hline
 Galaxy & $cz$ & $D_{L~K}$ & $D_{L~K,E}$ & ${H_0}_{~K}$ &
${H_0}_{~K,E}$ \\
  	& ${\rm km~s^{-1}}$ & Mpc & Mpc &
${\rm~km~s{}^{-1}~Mpc{}^{-1}}$ & ${\rm~km~s{}^{-1}~Mpc{}^{-1}}$ \\
\hline
($J$-band)\\
  PG0157 &  48869 &   251$\pm$  37 &   271$\pm$  40 &   202$\pm$  30 &   187$\pm$  27 \\
   IR02185 &  29347 &   421$\pm$ 100 &   441$\pm$ 105 &    71$\pm$  17 &    68$\pm$  16 \\
     IZW23 &  12224 &    40$\pm$   6 &    40$\pm$   6 &   313$\pm$  49 &   306$\pm$  48 \\
    CG1113 &  13880 &   118$\pm$  20 &   121$\pm$  21 &   118$\pm$  20 &   116$\pm$  20 \\
    IC2846 &  12294 &   208$\pm$  50 &   213$\pm$  51 &    60$\pm$  14 &    58$\pm$  14 \\
   IR14060 &  35060 &   562$\pm$  95 &   594$\pm$ 100 &    64$\pm$  11 &    61$\pm$  10 \\
    CG1417 &  21465 &   244$\pm$  44 &   253$\pm$  45 &    90$\pm$  16 &    86$\pm$  16 \\
    CG1448 &  13700 &   284$\pm$  53 &   291$\pm$  55 &    49$\pm$   9 &    48$\pm$   9 \\
     N6007 &  10547 &   177$\pm$  30 &   181$\pm$  30 &    60$\pm$  10 &    59$\pm$  10 \\
   IR16533 &  31808 &   526$\pm$ 156 &   561$\pm$ 167 &    62$\pm$  18 &    58$\pm$  17 \\
   IR23420 &  26022 &   493$\pm$  83 &   516$\pm$  87 &    54$\pm$   9 &    52$\pm$   9 \\
\\
($H$-band)\\
 PG0157 &  48869 &   235$\pm$  33 &   253$\pm$  36 &   216$\pm$  31 &   201$\pm$  28 \\
   IR02185 &  29347 &   421$\pm$  97 &   441$\pm$ 102 &    71$\pm$  16 &    68$\pm$  16 \\
     IZW23 &  12224 &    50$\pm$   8 &    51$\pm$   8 &   245$\pm$  37 &   240$\pm$  37 \\
    CG1113 &  13880 &   107$\pm$  18 &   110$\pm$  18 &   131$\pm$  22 &   128$\pm$  21 \\
    IC2846 &  12294 &   189$\pm$  44 &   193$\pm$  45 &    66$\pm$  15 &    64$\pm$  15 \\
   IR14060 &  35060 &   595$\pm$  98 &   629$\pm$ 103 &    61$\pm$  10 &    57$\pm$   9 \\
    CG1417 &  21465 &   260$\pm$  45 &   269$\pm$  47 &    84$\pm$  15 &    81$\pm$  14 \\
    CG1448 &  13700 &   369$\pm$  67 &   378$\pm$  69 &    38$\pm$   7 &    37$\pm$   7 \\
     N6007 &  10547 &   209$\pm$  34 &   213$\pm$  35 &    51$\pm$   8 &    50$\pm$   8 \\
   IR16533 &  31808 &   530$\pm$ 153 &   558$\pm$ 161 &    62$\pm$  18 &    58$\pm$  17 \\
   IR23420 &  26022 &   498$\pm$  81 &   519$\pm$  85 &    53$\pm$   9 &    51$\pm$   8 \\[4pt]
\hline
\end{tabular*}
\vskip 2mm
\noindent
Col.(1): Galaxy name. Col.(2): Redshift in $cz$.
Col.(3): Tully-Fisher distance corrected for $K$-correction.
Col.(4): Tully-Fisher distance corrected for $K$ and $E$-corrections.
Col.(5) and Col.(6): Hubble ratio for the Tully-Fisher distance in Col.(3) and Col.(4),
respectively.
\end{table*}

\newpage
%%%%%%%%%%%%%%%%%%%% Table 6 %%%%%%%%%%%%%%%%%
 \begin{table*}[t]
% \small
\begin{center}
Table~6. \hspace{4pt} Results of the CO-line Tully-Fisher relation.\\
\end{center}
\vspace{6pt}
\begin{tabular*}{\textwidth}{@{\hspace{\tabcolsep}
\extracolsep{\fill}}llc}
\hline\hline\\ [-6pt]
Band & Correction & $H_0$ \\
     &       & ${\rm km~s^{-1}~{Mpc}^{-1}}$\\[4pt]\hline\\[-6pt]
\\
$J$ 		& $K$	& ~60  $\pm$ 10 \\
$H$ 		& $K$	& ~53 $\pm$ 13 \\
\\
$J$ 		& $K+E$	& ~58  $\pm$ 10 \\
$H$ 		& $K+E$	& ~52  $\pm$ 12 \\[4pt]
\hline\\[-6pt]
\end{tabular*}

\vspace{6pt}\par\noindent
Col.(1): Used band.
Col.(2): Corrections applied to the total magnitude.
$K$ and $E$ stand for $K$-correction and $E$-correction,
respectively.
Col.(3): Determined Hubble constants.
\end{table*}

\end{document}